\documentclass{proceedingsM}
\usepackage{times,amsmath,amssymb}
\usepackage{graphicx,color}

\def\a{\alpha}
\def\b{\beta}

\def\A{\hat\alpha}
\def\B{\hat\beta}
\def\C{\hat\gamma}
\def\q{\tens q}
\def\Q{\tens Q}
\def\bm{\tens b}
\def\Bm{\tens B}
\def\j{\tens j}
\def\qI{\tens I}

\newcount\relabeling\relabeling=0  
\newcount\pageeqno\pageeqno=0  
\newcount\figures\figures=0  
\newcount\preview\preview=0    

\ifnum\preview=1  
\ifnum\preview=2  
  \AtBeginDocument{\Large}\makeatletter
  \def\section{\@startsection {section}{1}{\z@}{-3.5ex\@plus-1ex
    \@minus-.2ex}{2.3ex\@plus.2ex}{\normalfont\LARGE\bfseries}}
  \def\subsection{\@startsection{subsection}{2}{\z@}{-3.25ex\@plus-1ex
    \@minus-.2ex}{1.5ex\@plus.2ex}{\normalfont\Large\bfseries}}
  \makeatother\fi

\newcommand\nc{\newcommand*}  \nc\longnc{\newcommand}
                    \let\Z\kern  

\let\HB\hbox  \def\SB{\setbox1\HB}  \def\CB{\copy1}    
                \def\CC{\copy2}    
\def\RB#1{\raise#1\CB}    \def\YB{\ht1}  
\def\RC#1{\raise#1\CC}      
\newcount\n  \newdimen\w  \newdimen\h   
\def\mathsizes#1#2#3{\mathchoice{#1}{#1}{#2}{#3}}   

\def\tS#1{\ifcase#1\tiny\or          
  \scriptsize\or\footnotesize\or\small\or\normalsize\or\large\or
  \Large\or\LARGE\or\huge\or\Huge\else\ifnum#1<0\tiny\else\Huge\fi\fi}
\makeatletter                        
\def\cS{\ifx\@currsize\normalsize    
  4\else\ifx\@currsize\small 3\else\ifx\@currsize\footnotesize
  2\else\ifx\@currsize\large 5\else\ifx\@currsize\Large
  6\else\ifx\@currsize\LARGE 7\else\ifx\@currsize\scriptsize
  1\else\ifx\@currsize\tiny  0\else\ifx\@currsize\huge
  8\else\ifx\@currsize\Huge  9\else 4\fi\fi\fi\fi\fi\fi\fi\fi\fi\fi}
\DeclareRobustCommand\rS[1]{\ifmmode\@nomath\rS\else    
  \@tempcnta\cS\advance\@tempcnta#1\relax\tS\@tempcnta} 
\makeatother
\def\mS#1{\ifcase#1\displaystyle\or  
  \textstyle\or\scriptstyle\or\scriptscriptstyle\else\textstyle\fi}
\def\dS#1{\csname\ifcase#1relax\or   
  relax\or big\or Big\or bigg\or Bigg\fi\endcsname}

\def\textinmath#1{{\mathsizes                
  {\HB{#1}}{\HB{\tS1#1}}{\HB{\tS0#1}}}}      
\def\txt#1#2#3{\mskip#1mu                  
  \textinmath{#3}\mskip#2mu\relax }        

\def\re#1{(\ref{#1})}   

\long\def\quot#1{`#1'}    

\def\m#1{\scase=0$      #1         $}  
\def\mm#1{\m{       \,  #1  \,     }}  

\ifnum\pageeqno=1  
\usepackage{refcount}
\newcounter{lasteqpage}
\newcounter{lastlasteqpage}
\newcounter{lasteqno}
\newcounter{lastlasteqno}
\def\elabel#1{\label{#1}
  \setcounter{lastlasteqpage}{\thelasteqpage}
  \setcounter{lastlasteqno}{\thelasteqno}
  \setcounter{lasteqpage}{\getpagerefnumber{#1}}   
  \ifnum\thelasteqpage>\thepage\else\setcounter{lasteqpage}{\thepage}\fi
  \ifnum\thelasteqpage>\thelastlasteqpage\setcounter{lasteqno}{1}
  \else\addtocounter{lasteqno}{1}\fi}  
\def\non{\tagg\setcounter{lasteqpage}{\thelastlasteqpage}
              \setcounter{lasteqno}{\thelastlasteqno}}

\else  \def\elabel#1{\label{#1}}  \def\non{\tagg}  \fi

\makeatletter\@ifpackageloaded{amsmath}{
\def\eq#1#2{ \scase=1 \begin{align} \elabel{#1} #2 \end{align}}
\def\eqn#1#2{ \scase=1 \begin{align} \elabel{#1} \non #2 \end{align}}
\def\lel#1{ \\ \elabel{#1}     }  
\def\leln#1{\\ \elabel{#1} \non}  
\def\tagg{\tag*{}}  

  }{
\def\eq#1#2{  \scase=0\begin{equation}    \elabel{#1}   #2 \end{equation}}

\def\eqn#1#2{ \scase=0\begin{displaymath} \elabel{#1}   #2 \end{displaymath}}

\def\lel#1{\ifnum\scase=2\else\erroreqaneeded\fi \\ \elabel{#1}}
\def\leln#1{\ifnum\scase=2\else\erroreqaneeded\fi \\ \elabel{#1} \non}
\def\tagg{\nonumber}  
\def\lvert{|}  \def\rvert{|}  \def\lVert{\|}  \def\rVert{\|}
}\makeatother  

\newcount\scase  \def\7{&}  \def\s#1{\ifcase\scase#1\or\7#1\or\7#1\7\fi}

\def\0#1#2{\ifcase#1{#2}\or\lt(#2\rt)\or\lt[{#2}\rt]\or\lt\{{#2}\rt\}\or
  \mathord<{#2}\mathord>\or\lt\langle{#2}\rt\rangle\or\lt\lvert{#2}\rt
  \rvert\or\lt\lVert{#2}\rt\rVert\fi}
\def\1#1#2{\ifcase#1{#2}\or(#2)\or[#2]\or\{#2\}\or\mathord<{#2}\mathord
  >\or\langle{#2}\rangle\or\lvert{#2}\rvert\or\lVert{#2}\rVert\fi}
\def\2#1#2{\ifcase#1{#2}\or\big(#2\big)\or\big[#2\big]\or\big
  \{#2\big\}\or\big<#2\big>\or\big\langle#2\big\rangle\or\big
  \lvert#2\big\rvert\or\big\lVert#2\big\rVert\fi}
\def\3#1#2{\ifcase#1{#2}\or\Big(#2\Big)\or\Big[#2\Big]\or\Big\{#2\Big
  \}\or\Big<#2\Big>\or\Big\langle#2\Big\rangle\or\Big\lvert#2\Big
  \rvert\or\Big\lVert#2\Big\rVert\fi}
\def\4#1#2{\ifcase#1{#2}\or\bigg(#2\bigg)\or\bigg[#2\bigg]\or\bigg
  \{#2\bigg\}\or\bigg<#2\bigg>\or\bigg\langle#2\bigg\rangle\or\bigg
  \lvert#2\bigg\rvert\or\bigg\lVert#2\bigg\rVert\fi}
\def\5#1#2{\ifcase#1{#2}\or\Bigg(#2\Bigg)\or\Bigg[#2\Bigg]\or\Bigg
  \{#2\Bigg\}\or\Bigg<#2\Bigg>\or\Bigg\langle#2\Bigg\rangle\or\Bigg   
  \lvert#2\Bigg\rvert\or\Bigg\lVert#2\Bigg\rVert\fi}
\def\9#1#2{\ifcase#1{#2}\or\left(#2\right)\or\left[#2\right]\or\left
  \{#2\right\}\or\left\langle{#2}\right\rangle\or\left\langle{#2}\right
  \rangle\or\left\lvert{#2}\right\rvert\or\left\lVert{#2}\right\rVert\fi}
\def\lt{\mathopen{}\mathclose\bgroup\left} \def\rt{\aftergroup\egroup\right}

\let\f\frac                        
\def\fff{\largerfrac{-2}}        
\def\largerfrac#1#2#3{\mathchoice  
  {\SB{$\mS0\vcenter{}$}\w=\YB\SB{\rS{#1}$\mS0\vcenter{}$}
    \advance\w by-\YB\raise\w\HB{\rS{#1}$\mS0\frac{#2}{#3}$}}
  {\SB{$    \vcenter{}$}\w=\YB\SB{\rS{#1}$    \vcenter{}$}
    \advance\w by-\YB\raise\w\HB{\rS{#1}$    \frac{#2}{#3}$}}
  {\SB{$\mS2\vcenter{}$}\w=\YB\SB{\rS{#1}$\mS2\vcenter{}$}
    \advance\w by-\YB\raise\w\HB{\rS{#1}$\mS2\frac{#2}{#3}$}}
  {\SB{$\mS3\vcenter{}$}\w=\YB\SB{\rS{#1}$\mS3\vcenter{}$}
    \advance\w by-\YB\raise\w\HB{\rS{#1}$\mS3\frac{#2}{#3}$}}}

\def\tr{\mathop{\txt00{tr}}}

\def\dd{\mathrm{d}}    
\let\pd\partial

\def\tensUpRt#1{^{\mathrm{#1}}}            
   
\def\dev{\tensUpRt{d}}   \def\sph{\tensUpRt{s}}
\let\tens\mathbf  
\let\Tens\mathBf  

\def\paraline{\rule{.03em}{.9ex}}
\def\para{_{}^{\paraline\kern.06em\paraline}}  

\def\qirr#1{{\hat{#1}}}

\def\qcurr#1#2{{#1}_{#2}}  
\def\qini{}
\def\qext#1{{\tilde{#1}}}

\def\qe{e}
\def\qj{j}  \def\qqj{\tens{\qj}}
  
\def\ql{l}

\def\qs{s}

\def\qT{T}

\def\qalp{E_0}
\def\qbet{E_1}  \def\qgam{E_2}  \def\qtau{\tau}
\def\qeps{\varepsilon}
\def\qqeps{\Tens\qeps}
\def\qlam{\lambda}
\def\qrho{\varrho}
\def\qsig{\sigma}  \def\qqsig{\Tens{\qsig}}
\def\qxi{\xi}  \def\qqxi{\Tens{\qxi}}

\title{Thermodynamic hierarchies of evolution equations}  
\author{V\'an P.\footnote, Kov\'acs R.\footnote and F\"ul\"op T.\footnote}
\address{{\footnotemark[1]} Department of Theoretical Physics, Wigner Research 
Centre for Physics, Institute for Particle and Nuclear Physics, Budapest, 
Hungary}
\address{{\footnotemark[2]} Department of Energy Engineering, BME, Budapest, 
Hungary}
\address{{\footnotemark[3]}Montavid Thermodynamic Research Group} 

\author{V\'an P.$^{abc}$, Kov\'acs R.$^{abc}$ and Ful\"op T.$^{bc}$}
\address{$^{a}$ Department of Theoretical Physics, Wigner Research 
Centre for Physics,
Institute for Particle and Nuclear Physics, Budapest, Hungary}
\address{$^{b}$  Department of Energy Engineering, BME, Budapest, 
Hungary}
\address{$^{c}$ Montavid Thermodynamic Research Group}
\email{van.peter@wigner.mta.hu}

\abstract{Non-equilibrium thermodynamics with internal variables introduces a 
natural hierarchical arrangement of evolution equations. Three examples are 
shown: a hierarchy of linear constitutive equations in thermodynamic rhelogy 
with a single internal variable, a hierarchy of wave 
equations in the theory of generalized continua with dual internal 
variables and a hierarchical arrangement of the Fourier equation in 
the theory of heat conduction with current multipliers.
}

\keywords{multiscale, non-equilibrium thermodynamics, internal variables} 
\begin{document}
\maketitle

\section{Introduction}
\label{intro}

Theories and material models of multiscale phenomena in space and 
time treat the scale changes either as a step from a micro- or mesoscopic 
statistical level to the phenomenological one or as a reduction of the degrees 
of freedom by averaging over a field variable or spatial dimension. The 
characteristic methodology is similar to the BBKGY hierarchy of the kinetic 
theory \cite{GrmEta11a,Lib90b}. In these approaches, the modeling of 
transitional effects requires the detailed knowledge or the drastic 
simplification of the material structure. 

In this work, we show that a hierarchical arrangement of evolution equations
is apparent in thermodynamics with internal variables where the different 
levels of the hierarchy are regulated by material parameters. The scale 
transitions are natural and dynamical. 

We show three examples. First, a hierarchy of ordinary differential equations is 
presented in the  thermodynamic rheology of solids with a single internal 
variable. The building block of the hierarchy is the basic constitutive 
equation of elasticity. This is a hierarchy between the different 
time scales of the evolution, a {\em time hierarchy}. 

Then a hierarchy of hyperbolic partial differential  equations is shown in the 
thermodynamic theory of generalized continua with dual internal variables. The 
building block of the hierarchy is the wave equation. This is a dynamic 
hierarchy between different time and length scales of the evolution, a {\em 
space-time hierarchy}.

Finally, a hierarchy of parabolic partial differential  equations is shown in 
the thermodynamic theory of heat conduction with current multipliers. The 
building block of the hierarchy is the Fourier equation. Like the previous 
example, it is a dynamic hierarchy between different time and length scales of 
the evolution, a {\em space-time hierarchy}.

\section{The hierarchy of rheological bodies and the Kluitenberg--Verh\'as model}

In the thermodynamic approach to rheology, an extended state space is chosen, 
which is spanned by the following variables: specific internal energy \m 
{\qini\qe }, strain \m {\qqeps}, and an internal variable \m{ \qqxi }. This 
modeling approach is well-known for fluids 
\cite{Klu62a1,Ver97b,KluCia78a,CiaKlu79a,MauMus94a1} and has been
introduced recently for solids \cite{AssEta14a}. The internal variable is a 
second order symmetric tensor, based on our purpose to gain an extension
of the mechanical aspects \1 1 {the \quot{material law}} of the initial
system, to obtain corrections to the relation between stress and strain,
which quantities are both symmetric tensors.

We shift entropy by a concave nonequilibrium term
depending---quadratically---on \m { \qqxi } only. According to the Morse
lemma, this new entropy term can be chosen as a pure square term, hence,
the extended specific entropy function $\qext{\qs}$ is
 \eq{.1.1.}{
\qext{\qs} \1 1 {\qini\qe, \qqeps, \qqxi} =
\qini\qs \1 1 {\qini\qe, \qqeps} - \f {1}{2} \tr \0 1 { \qqxi^2 },
 }
where $\tr$ denotes the trace of a second order tensor and we have denoted the 
classical specific entropy without tilde. The Gibbs relation for the extended 
entropy is a convenient particular thermodynamic notation for the partial 
derivatives, the intensive quantities:
 \eq{.1.2.}{
\qrho \dd \qext\qs = \f{\qrho}{\qT} \dd \qini\qe - \f{1}{\qT} \tr \1 1
{\qini\qqsig \dd \qqeps} - \qrho \tr \0 1 { \qqxi \dd \qqxi } .
 }
Here $\qrho$ is the density, $\qT$ is the temperature and $\qqsig$ is the 
thermostatic stress. Stress is also
considered extended by a rheological \1 1 {nonequilibrium} term:
 \eq{.1.3.}{
\qext{\qqsig} = \qini{\qqsig} + \qirr{\qqsig} .
 }
Consequently, the mechanical power, and correspondingly the
energy balance, gets shifted as
 \eq{.1.4.}{
\qrho \dot{\qini\qe} + \nabla \cdot \qcurr{\qqj}{\qini\qe} = \tr \1 1 {
\qext\qqsig \dot \qqeps } = \tr \1 1 { \qini\qqsig \dot \qqeps } + \tr
\1 1 { \qirr\qqsig \dot \qqeps } .
 }
Here $\qcurr{\qqj}{\qini\qe}$ is conductive current density of the internal 
energy, the heat flux.  With the choice \mm {\qcurr{\qqj}{\qext\qs} = 
{\qqj_{\qini\qe}}/{\qT},} and utilizing \re{.1.2.} and \re{.1.4.}, the entropy 
production is found to be
 \eqn{@136.}{
\Sigma \s= \qrho \dot{\qext\qs} + \nabla \cdot
\qcurr{\qqj}{\qext\qs} = \f{\qrho}{\qT} \dot{\qini\qe} - \f{1}{\qT} \tr
\1 1 {\qqsig \dot{\qqeps}} - \qrho \tr \2 1 { \qqxi \dot \qqxi } +
\nabla \cdot \9 1 { \f{\qcurr{\qqj}{\qini\qe}}{\qT} }
 \leln{.1.5.}
\s= - \f {1}{\qT} \nabla \cdot \qcurr{\qqj}{\qini\qe} + \f {1}{\qT} \tr
\1 1 { \qirr\qqsig \dot \qqeps } - \qrho \tr \2 1 { \qqxi \dot \qqxi }
+ \nabla \cdot \9 1 { \f{\qcurr{\qqj}{\qini\qe}}{\qT} }
 \lel{.1.6.}
\s= \qcurr{\qqj}{\qini\qe} \cdot \nabla  \0 1 { \f {1}{\qT} } +
 \f{1}{\qT} \tr \2 1 { \qirr\qqsig\dev \dot \qqeps\dev }
+ \f{1}{\qT} \tr \2 1 { \qirr\qqsig\sph \dot \qqeps\sph }
- \qrho \tr \2 1 { \qqxi\dev \dot \qqxi{}\dev }
- \qrho \tr \2 1 { \qqxi\sph \dot \qqxi{}\sph } \geq 0.
 }
In the rhs, vectors are present in the first term, scalars in the third
and fifth one, and symmetric traceless tensors in the second and fourth
term. In an isotropic material, these three types of quantities cannot
couple to one another. Therefore, concerning the term containing
vectors, we consider Fourier heat conduction, \mm { \qcurr{\qqj}{\qe} =
\qlam \nabla \0 1 { \f {1}{\qT} } }. For the remaining two pairs of
terms, the most general Onsagerian solution is
 \eq{.1.7.}{
 \qquad
\qirr\qqsig\dev \s= \ql_{11}\dev \dot \qqeps\dev +
\ql_{12}\dev \2 1 {\mathord- \qrho \qT \qqxi\dev} ,
 &
\qirr\qqsig\sph \s= \ql_{11}\sph \dot \qqeps\sph +
\ql_{12}\sph \2 1 {\mathord- \qrho \qT \qqxi\sph} ,
 \qquad
 \leln{.1.8.}
 \qquad
\dot \qqxi{}\dev \s= \ql_{21}\dev \dot \qqeps\dev +
\ql_{22}\dev \2 1 {\mathord- \qrho \qT \qqxi\dev} ,
 &
\dot \qqxi{}\sph \s= \ql_{21}\sph \dot \qqeps\sph +
\ql_{22}\sph \2 1 {\mathord- \qrho \qT \qqxi\sph} ,
 }
where the  $\ql_{11}\dev,\ql_{12}\dev,\ql_{21}\dev,\ql_{22}\dev$ and  
$\ql_{11}\sph ,\ql_{12}\sph ,\ql_{21}\sph ,\ql_{22}\sph $ material parameters 
are subjects of thermodynamic restrictions, due to the entropy inequality 
\re{.1.6.}. 

Eliminating the internal variable in the constant temperature case also
leads to two independent models,
 \eq{.1.9.}{
\qqsig\dev + \qtau\dev \dot\qqsig\dev \s = \qalp\dev \qqeps\dev +
\qbet\dev \dot \qqeps\dev + \qgam\dev \ddot \qqeps\dev ,
 &
\qqsig\sph + \qtau\sph \dot\qqsig\sph \s = \qalp\sph \qqeps\sph +
\qbet\sph \dot \qqeps\sph + \qgam\sph \ddot \qqeps\sph ,
 }
with thermodynamics-originated inequalities for the altogether eight
coefficients. The complete model is a deviatoric and a spherical {\em 
Kluitenberg--Verh\'as body}. When $\qgam\dev=0$, the deviatoric part reduces to 
the standard or Poynting--Thomson body of solid rheology. Several simpler 
rheological bodies may be obtained by a particular choice of the parameters.

A suitable rearrangement reveals the hierarchical structure of the equations: 
 \eqn{.1.10.}{
(\qqsig\dev-\qalp\dev \qqeps\dev) + 
\qtau\dev \f{\dd}{\dd t} \left( \qqsig\dev
 - \f{\qbet\dev}{\qtau\dev} 
\qqeps\dev \right) +
\qgam\dev \f{\dd^2}{\dd t^2} \qqeps\dev = 0 ,
 \\
(\qqsig\sph-\qalp\sph \qqeps\sph) + 
\qtau\sph \f{\dd}{\dd t}\left(\qqsig\sph
 - \f{\qbet\sph}{\qtau\sph} 
\qqeps\sph \right) +
\qgam\sph \f{\dd^2}{\dd t^2} \qqeps\sph = 0 ,
 }
In both the deviatoric and the spherical cases, the first term is the pure 
elastic stress--strain relation, the 
second is the time derivative of a similar relation with different coefficients 
and the third one with the highest derivative is an incomplete block, closing 
the two terms' hierarchy. 

If the coefficients in the consequtive blocks are the same, we may speak about 
{\em hierarchical resonance}. If the closure term is zero, a hierarchical 
resonance may be not dissipative. 

In case of specific loading conditions, the deviatoric and spherical parts are 
coupled but the hierarchical structure may be conserved. It is straightforward 
to calculate the effective rheological equation in case of uniaxial loading 
conditions. Denoting the uniaxial stress by $\qsig\para$, one
obtains: 
\eq{.1.11.}{
\qsig\para -  {E\para}_0\qeps\para +
{\qtau\para}_1 \f{\dd}{\dd t}\left(\qsig\para -  \f{{E\para}_1}{{\qtau\para}_1} 
	\qeps\para \right)+
{\qtau\para}_2 \f{\dd^2}{\dd t^2}\left(\qsig\para - \f{{E\para}_2}
	{{\qtau\para}_2} \qeps\para \right)+
{\qtau\para}_3 \f{\dd^3}{\dd t^3}\left(\qsig\para -  \f{{E\para}_3}
	{{\qtau\para}_3} 	\qeps\para \right)+
{E\para}_4\f{\dd^4}{\dd t^4}\qeps\para = 0
}

Where ${\qtau\para}_1, {\qtau\para}_2, {\qtau\para}_2$ and 
${E\para}_0,{E\para}_1,{E\para}_2,{E\para}_3$ coefficients are calculated from 
the spherical and deviatoric coefficients of \re{.1.9.} \cite{AssEta14a}.

In typical experimental situations, the time scales of the different blocks are 
clearly separated. 

\section{Hierarchy of wave equations in the  theory of dual internal variables}

Dual internal variables extend the modeling capability of non-equilibrium 
thermodynamics by connecting inertial phenomena with dissipation 
\cite{VanAta08a}. Dual internal variables coupled to continuum mechanics lead 
to generalized continua \cite{BerEta11a1,VanEta14a}. In this case, the 
elimination of the internal variables results in a hierarchical structure of 
wave equations \cite{BerEta10a2,BerEta11a1,BerEng13a}. 

In
what follows,
 we introduce
in brief a one dimensional version of the theory 
of weakly nonlocal dual internal variables coupled to small-strain elasticity. 
Therefore, the extended state space is given by the strain, $\qeps$, and the  
internal variables are denoted by $\phi$ and $\xi$. In this illustrative 
example, specific entropy is a quadratic function of the internal variables 
and their gradients:
\eq{.2.1.}{
\tilde s(e,\qeps,{ \xi}, 	{\phi, 
	\partial_x\xi, \partial_x \phi}) = 
s(e,\qeps) - {
	\f{a_1}{2}\xi^2}-	
	\f{b_1}{2}(\partial_x\xi)^2-
	\f{a_2}{2}\phi^2- 	
	\f{b_2}{2}(\partial_x\phi)^2. 
}
The Gibbs relation of the weakly nonlocal theory fixes the partial derivatives 
of the entropy function as 
\eq{.2.2.}{
\rho \dd\tilde s = \f{\rho}{T} \dd e - \f{\sigma}{T} \dd\qeps - 
	\rho a_1 \xi\ \dd\xi - 
	\rho a_2 \phi \dd\phi -
	\rho b_1 \partial_x\xi \dd (\partial_x \xi)- 
	\rho b_2 \partial_x\phi \dd (\partial_x \phi)
}
Assuming the following form of the entropy current density:
\eq{.2.3.}{
j_{\tilde s} = \f{q}{T} -
	\rho \f{\pd s}{\partial (\partial_x \xi)} \dot \xi -
	\rho \f{\pd s}{\partial (\partial_x \phi)} \dot \phi ,
}
one obtains the entropy production similarly to the previous section:
\eq{.2.4.}{
T\Sigma = T q \pd_x\f{1}{T} +
\left(\sigma - T \rho \partial_\qeps s\right) \dot \qeps + 
	\underbrace{\big(\pd_\xi s - \pd_x (\pd_{\pd_x\xi} s) 
		\big)}_{\hat A \xi}\dot \xi+
	\underbrace{\big(\pd_\phi s - \pd_x (\pd_{\pd_x\phi} s) 
	\big)}_{\hat B\phi}\dot \phi \geq 0 .
}
Here, we have introduced a shorthand notation for the internal variable related 
weakly nonlocal thermodynamic forces. The above form of the entropy current 
density and entropy production (dissipation inequality) can be also derived 
with the help of a more detailed thermodynamic analysis, as it has been shown in 
\cite{VanEta14a}. Then a linear solution of the above inequality is
\eqn{.2.5.}{
\sigma- E\qeps  &= l_{11}\dot\qeps +
	l_{12}\hat A \xi + l_{13}\hat B \phi , \leln{2.52}
\dot \xi &= l_{21}\dot\qeps +
	l_{22}\hat A \xi + l_{23}\hat B \phi ,\lel{2.5}
\dot \phi &= l_{31}\dot\qeps +
		l_{32}\hat A \xi + l_{33}\hat B \phi. 
}
In our simple case 
\eq{.2.6.}{
\hat A = - a_1 + b_1 \pd_{xx}, \qquad
\hat B = - a_2 + b_2 \pd_{xx}.
}
The elimination of the internal variables leads to the following constitutive 
relation of stress and strain:
\eqn{.2.7.}{
&\ddot \sigma + 
(\a_1 + \a_2 \pd_x)\dot \sigma +
(\b_1 + \b_2 \pd_x + \b_3 \pd_{xx})
 \sigma  = \lel{2.7}
 & \qquad\dddot \qeps + 
(\A_1 + \A_2 \pd_x)\ddot \qeps +
(\B_1 + \B_2 \pd_x + \B_3 \pd_{xx})\dot \qeps +
(\C_1 + \C_2 \pd_x + \C_3 \pd_{xx}) \qeps, 
}
where the
 coefficients
 $\a_1,\a_2,\b_1,\b_2,\b_3$ and 
$\A_1,\A_2,\B_1,\B_2,\B_3,\C_1,\C_2,\C_3$
 are simple polinomials 
of the thermodynamic material parameters. The consequence of the momentum 
balance and the compatibility condition leads 
to the well-known relation of
 stress and strain
\eq{.2.8.}{
\rho \dot v - \pd_x \sigma = 0, \qquad 
\pd_x v = l_1\dot \qeps \quad
 \Rightarrow
 \quad \ddot\qeps = \pd_{xx} \sigma.
}

Eliminating
 stress from  \re{2.7}, one obtains the following partial 
differential equation:
\eqn{.2.9.}{
&(\ddot \qeps - \A_1 \pd_{xx}\qeps\ddot)+ 
(\a_1 \ddot \qeps - \B_1 \pd_{xx}\qeps\dot )+ 
(\b_1 \ddot \qeps - \C_1 \pd_{xx}\qeps)+ \\&\qquad
\pd_x(\b_2 \ddot \qeps - \C_2 \pd_{xx}  \qeps\dot)+ 
\pd_{xx}(\b_3 \ddot \qeps - \C_3 \pd_{xx} \qeps) -
\pd_{xx}(l_1 \ddot \qeps + \B_3 \pd_{xx}\qeps + \a_2 \pd_{x}\dot \qeps\dot)=0.
}
Mixed space and time derivatives of coupled wave equations are 
analysed in detail and are compared to
various
 wave propagation models in 
\cite{BerEng13a}.

\section{Hierarchy of Fourier equations and generalized heat conduction with 
current multipliers}

Non-equilibrium thermodynamics with current multipliers introduces a unified 
constitutive theory of heat conduction where several generalizations of Fourier 
equation may be obtained as special cases \cite{VanFul12a}. Moreover, it is 
shown that the structure is compatible with
 the
moment series expansion of kinetic 
theory, at least up to the third moment \cite{KovVan14m}. In this
framework, the 
basic state space is extended by the heat flux $\q$ and also by a second order 
tensorial internal variable $\Q$. We assume the usual quadratic form of the 
entropy function at the extended part of the state space,
\eq{.4.1.}{
  \tilde s(e, \q,\Q)=s(e)-\frac{m_1}{2} \q^2 - \frac{m_2}{2} \Q^{2}.
}

Then a generalized entropy current is introduced in the following form:
\eq{.4.2.}{
\j_q = \bm \cdot \q + \Bm : \Q .
}
Here, the
 current multipliers
 $\bm$ and $\Bm$
 are second and third order tensors, 
respectively. This form of the generalized entropy current was introduced by 
Ny\'iri \cite{Nyi91a1}. $\bm$ and $\Bm$ are to be determined as constitutive 
functions with the help of the entropy inequality. A short calculation results 
in 
\eq{.4.3.}{
\Sigma = \left(\bm - \frac{1}{T}\qI \right): \nabla \q - 
\left( \nabla \bm -  m_1 \dot \q\right)\cdot {\q} - 
\left( \nabla\cdot \Bm -  m_2 \dot \Q\right): {\Q} + 
\Bm \cdot : \nabla \Q \geq 0.
}
Here the number of the central dots denotes one, two and three contractions of 
the first, second and third order tensors, respectively. The first and the 
third terms are products of second order tensors, the second 
term is vectorial, and the last term is a product of third order tensors. 
Therefore,
for isotropic materials,
 cross effects may appear only between the first and the third terms. Hence, in a
one dimensional simplification, linear 
relations between the {thermodynamic fluxes} and {forces} are as
follows:
\eq{.4.4.}{
m_1 \dot q -\pd_x b &= -l_1 {q}, \lel{4.41}
m_2 \dot Q -  \pd_x B &= -k_1 {Q} + k_{12} \pd_x q, \lel{4.42}
b- \frac{1}{T} &= -k_{21} {Q} + k_{2}\pd_xq, \lel{4.4}
B &= n_3 \pd_x Q, 
}
where $\pd_x$ denotes the one dimensional spatial derivative and the material 
coefficients $m_1,m_2,l_1,k_1,k_2,k_{12},k_{21},n_3 $ are subjects
to 
thermodynamical constraints. It is straightforward to eliminate the current 
multipliers and the  tensorial 
internal variable $Q$. Then one obtains the following equation:
\eqn{.4.5.}{
m_1 m_2 \ddot q &+ (m_2 l_1 +m_1 k_1)\dot q  - (m_2k_2+m_1 n_3) \pd_{xx}\dot q +
k_1l_1 q  - (k_1k_2-k_{12}k_{21} +l_1n_3)  \pd_{xx}q + k_2n_3 \pd_{xxxx}q -
 \lel{.4.5b.}
&k_1  \pd_{xt}\left(\f{1}{T} \right)-m_2\pd_{x}\left(\f{1}{T} \right)
+m_3\pd_{xxx}\left(\f{1}{T} \right) = 0
}
In our case, the balance of internal energy \re{.1.4.} is 
\eq{.4.6.}{
\rho c \dot T +\pd_{x}q = 0,
}
where $\rho$ is the density and $c$ is the specific heat.
The combination of \re{.4.5b.} and \re{.4.6.} may be written in the following 
form:
\eqn{.4.7.}{
&\Big( (m_2l_1 + m_1 k_1 ) \dot T - 
	\fff{m_2}{\rho c T^2}\pd_{xx}{T} \ddot{\Big)} + 
k_1\Big( l_1 \dot T - \fff{1}{\rho c T^2}\pd_{xx}{T} \dot{\Big)} + 
\leln{4.71}&\qquad
\pd_{xx}\left( (k_1k_2-k_{12}k_{21} +l_1n_3) \dot T - 
	\fff{n_3}{\rho c T^2}\pd_{xx}{T} \right) + 
\lel{4.7}&\qquad
\big(m_2 m_1 \dot T - (m_1n_3+m_2k_2)\pd_{xx}T \ddot{\big)}  + 
k_2n_3 \pd_{xxxx}\dot T = 0.
}
We
can observe various time and space derivatives of the Fourier equation in 
different forms, plus the last term with the highest derivatives.  The 
arrangement is space-time hierarchical, like in the previous section.

\subsection{Hierarchical resonance: the example of the
Guyer--Krumhansl equation}

The hierarchical rearrangement of an evolution equation may help in recognizing 
solution patterns. In this subsection, we give a simple example with the
help of the Guyer--Krumhansl equation.

The Guyer--Krumhansl equation is obtained when $n_3=k_2=m_1=k_{12}=0$ in 
\re{4.7}:
\eq{.5.1.}{
\tau_q \pd_t\left(\pd_{t} T - 
\f{a}{\tau_q}\pd_{xx} T\right) + 
\pd_{t} T - \lambda\pd_{xx} T=0.
}
Here, $\tau_q= m_2/k_1$, $\lambda = 1/(\rho c k_1l_1 T^2)$ and $a=m_2/(\rho c 
k_1l_1)$, and these coefficients are considered constant. This is a two-level 
hierarchical arrangement. If $\tau_q=0$ then the first term is 
zero, and the hierarchy is reduced to a single Fourier equation. If $a= \tau_q 
\lambda$ then there appears the same Fourier equation in both 
terms. This is the case of {\em hierarchical resonance} \cite{KovVan14m} and 
the solutions of the coupled set of equations may be identical to the single  
Fourier equation. 

The resonance may help to classify the solutions. Let us introduce adiabatic 
boundary at the end of a rod and heat pulse boundary conditions at the the 
front side in the following form:
\begin{center}
$q_0(t)=q(x=0,t)= \left\{ \begin{array}{cc}
q_{\rm max}\left(1-\cos\left(2 \pi \cdot \frac{t}{t_p}\right)\right) & 
\textrm{if}\;\; 0<t \leq t_p,\\
0 & \textrm{if}\;\; t>t_p.
\end{array} \right.  $
\end{center}
Here $t_P$ is the duration of the pulse and $q_{\rm max}$ is the maximum of the 
heat flux at the boundary. Initially, the temperature is uniform and there is no 
heat flux $q(t=0,x)=0$, $T(t=0,x)=T_0$. Then the solutions show characteristic 
differences depending on whether the parameters are above or below the resonance 
value. This is represented on Figure 1, where temperature and time are 
the following dimensionless quantities: $\hat{t} =\frac{\lambda t}{\rho c L^2}$ 
and $\hat{T} =\frac{T-T_{0}}{T_{\rm end}-T_{0}}$. Here $L$ is the length of the rod 
and $T_{\rm end}$ is the asymptotic value of the temperature after the 
equilibration. 
\begin{itemize}
\item If  $a= \tau_q \lambda$ then we obtain the solution of the Fourier 
equation. This is the solid line on Fig. 1. 
\item If  $a< \tau_q \lambda$ then we obtain solutions where temperature 
starts to increase later than in the Fourier solution. For short rods, the 
heat pulse is observable. The important characteristics of the solution are 
similar the solutions of the Maxwell--Cattaneo--Vernotte equation. This is  
the dashed line on Fig. 1.
\item If  $a>\tau_q \lambda$ then temperature starts to increase
earlier 
than
for the Fourier solution. The remnants of the heat pulse are not 
observable, and sometimes there is a change in the steepness of the solution, a 
kink. The solution is
more
 damped 
 than the Fourier one. This is  the dashed-dotted line on Fig. 1.
\end{itemize}
\begin{figure}[ht]
\centering
\includegraphics[width=10cm,height=7cm]{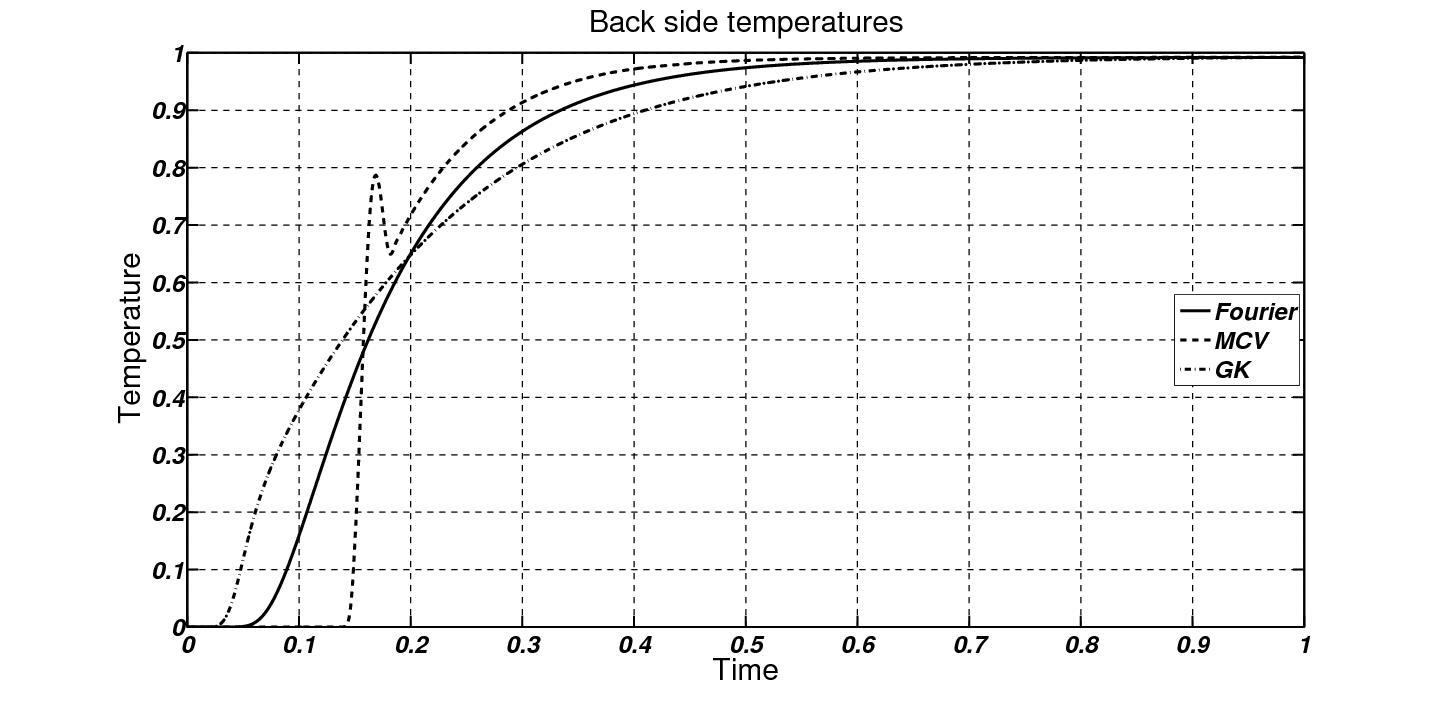}  
\caption{Characteristic solutions of of the Guyer--Krumhansl equation in case of 
heat pulse experiments.}
\label{fig:allin}
\end{figure}

\section{Summary}

Eliminating internal variables in non-equilibrium thermodynamics results
in a hierarchical structure of the evolution equation. The buiding block
of the hierarchy is the evolution equation of the original theory, which
was supplemented by the internal variable.

The solution of the original equation may appear at different particular values 
of the parameters. If this happens with more than one nonzero elements of the 
hierarchy, we can  call it hierarchical resonance.

\section{Acknowledgement}  

The work was supported by the grants OTKA K81161 and K104260. 


\end{document}